\documentclass[10pt,a4paper,english,nofootinbib,superscriptaddress,aps,preprint]{revtex4-1}
\usepackage[T1]{fontenc}
\usepackage[utf8]{inputenc}
\usepackage{amsmath}
\usepackage{amssymb}
\usepackage{esint}
\usepackage{graphicx}

\makeatletter

\makeatletter\usepackage{babel}

\usepackage{hyperref}

\makeatother

\begin{document}

\title{Noncommutative Space Corrections on the Klein-Gordon and Dirac Oscillators Spectra}

\author{Roberto V. Maluf}
\affiliation{Instituto de F\'{\i}sica, Universidade de S\~ao Paulo\\Caixa Postal 66318, 05315-970, S\~ao Paulo, SP, Brazil}
\email{rmaluf@fma.if.usp.br}

\begin{abstract}
We consider the influence of a noncommutative space on the Klein-Gordon
and the Dirac oscillators. The nonrelativistic limit is taken and
the $\theta$-modified Hamiltonians are determined. The corrections
of these Hamiltonians on the energy levels are evaluated in first-order
perturbation theory. It is observed a total lifting of the degeneracy
to the considered levels. Such effects are similar to the Zeeman splitting
in a commutative space.
\end{abstract}

\maketitle

\section{Introduction}

The idea of a noncommutative space-time has been under intense investigation in recent years, since its resurgence in connection with string theory \cite{Witten}. The subject has received a great deal of attention and many studies have been developed involving different respects (see e.g. Refs. \cite{RevQFT} and \cite{RevQM} for reviews on noncommutativity in quantum field theory and quantum mechanics, respectively).   

In the context of quantum mechanics, the noncommutative space can be implemented by the coordinate operators $\hat{x}^{i}$ and the conjugate momenta $\hat{p}^{i}$, satisfying commutation relations:
\begin{equation}
[\hat{x}^{i},\hat{x}^{j}]=i\theta^{ij}, \ \ [\hat{x}^{i},\hat{p}^{j}]=i\hbar \delta^{ij}, \ \ [\hat{p}^{i},\hat{p}^{j}]=0,
\label{eq.1}\end{equation}
where $\theta^{ij}$ is  a real and antisymmetric parameter matrix with dimension of length
squared. It is shown in Ref. \cite{Gamboa} that the relation between noncommutative and commutative variables can be obtained by a linear transformation:
\begin{equation}
\hat{x}^{i}=x^{i}-\frac{1}{2\hbar}\theta^{ij}p^{j}, \hspace{1cm} \hat{p}^{i}=p^{i},
\end{equation} with $x^{i}$ and $p^{i}$ satisfying the commutation relations as the usual commutative space.

A well-known result involving the spatial noncommutativity is the splitting on the energy spectrum of the hydrogen atom  \cite{Chaichian,Thiago}. 
Other interesting result is the connection of the chiral oscillator with the noncommutative space, as was shown in Ref. \cite{Banerjee}. In that work, the chiral oscillator is determined from the usual harmonic oscillator such that the Poisson brackets between the coordinates exhibit a similar structure to that in Eq. (\ref{eq.1}). Some applications of the chiral oscillator, such as its connection with the electric-magnetic duality and the Zeeman effect  were then reported.

In another paper \cite{Mirza}, the noncommutative versions of the Klein-Gordon and Dirac oscillators were initially discussed. In both cases, the relativistic equations of motion in a noncommutative space were determined and their similarities with those of a particle in a commutative space with a constant magnetic field have been reported. As claimed by the authors, in the case of a particle with spin the problem is not exactly soluble in three spatial dimensions. This fact, in the context of a perturbative treatment, opens the possibility of investigating the issue concerning the nonrelativistic corrections, induced by spatial noncommutativity on the energy spectrum of the system under consideration. Furthermore, it is interesting to answer the question whether such corrections are able to completely lift the degeneracy of the energy levels.

The present work has as its main goal to examine the noncommutativity effects on the Klein-Gordon and Dirac oscillators, with special emphasis to the nonrelativistic limit and possible modifications on the energy levels. Our calculations are performed in the framework of the degenerate perturbation theory, by considering the $\theta$-modification as a first-order perturbation. We show that in both cases, the spatial noncommutativity is able to completely remove the degeneracy of the levels analyzed. This correction is similar to the Zeeman effect and proportional to the $\theta$-parameter magnitude. 

Our paper is organized as follows. In Sec. 2, we study the noncommutativity effects on the nonrelativistic limit of the Klein-Gordon oscillator. The corrections to the energy levels are evaluated by first-order perturbation theory. In Sec. 3, we extend our investigation to the Dirac oscillator. In Sec. 4, we present our conclusions.


\section{Energy Corrections of the Klein-Gordon Oscillator in a Noncommutative
Space}

The equation that describes the behaviour of a relativistic spin-zero
particles is the well-known Klein-Gordon equation. For a free particle
in 4-dimensional space-time it reads $(\square+m^{2}c^{2}/\hslash^{2})\varphi=0,$
where $m$ is the rest mass of the particle and $c$ the velocity
of light. The Klein-Gordon oscillator can be obtained through the
following non-minimal substitution in the free equation, \begin{equation}
c^{2}\left(\mathbf{p}+im\omega\mathbf{x}\right)\cdot\left(\mathbf{p}-im\omega\mathbf{x}\right)\varphi(\mathbf{x})=\left(W^{2}-m^{2}c^{4}\right)\varphi(\mathbf{x}),\label{eq:Klein1}\end{equation}
where $\mathbf{p}=-i\nabla$, $\omega$ is the oscillator frequency
and we take the time dependence of the wave function to be $\varphi(\mathbf{x},t)=\varphi(\mathbf{x})\exp(-iWt/\hslash)$. 

Now, let us consider the space noncommutativity version of the equation above. As shown in Ref. \cite{Mirza}, it can be represented as
\begin{equation}
c^{2}\left[\mathbf{p}+im\omega\left(\mathbf{x}+\frac{\boldsymbol{\theta}\times\mathbf{p}}{2\hslash}\right)\right]\cdot\left[\mathbf{p}-im\omega\left(\mathbf{x+\frac{\boldsymbol{\theta}\times\mathbf{p}}{2\hslash}}\right)\right]\varphi=\left(W^{2}-m^{2}c^{4}\right)\varphi,\label{eq:Klein1.1}\end{equation}
where $\theta_{i}=\frac{1}{2}\varepsilon_{ijk}\theta_{jk}$ is the
constant noncommutativity parameter. Let us write Eq. (\ref{eq:Klein1.1}) more explicitly, \begin{equation}
c^{2}\left[p^{2}+m^{2}\omega^{2}r^{2}-3m\hbar\omega-\frac{m^{2}\omega^{2}}{\hbar}\boldsymbol{\theta}\cdot\mathbf{L}+\frac{m^{2}\omega^{2}}{4\hbar^{2}}(\boldsymbol{\theta}\times\mathbf{p})^{2}\right]\varphi=\left(W^{2}-m^{2}c^{4}\right)\varphi,\label{eq.2}\end{equation}
where $r=\sqrt{\mathbf{x}\cdot\mathbf{x}}$ and $\mathbf{L}=\mathbf{x}\times\mathbf{p}$ is the orbital angular
momentum of the particle.

To set out the form and physical meaning of the possible effects related
to the noncommutative space, we shall confine ourselves to the nonrelativistic
regime. In this case, the energy is concentrated mainly in the mass
of the particle and we can write $W=E+mc^{2}$. Therefore, in the nonrelativistic limit: $W^{2}-m^{2}c^{4}\approxeq 2mc^{2}E $ for $E \ll mc^{2}$ being the nonrelativistic energy. Taking this limit in Eq. (\ref{eq.2}) and dividing through by $2mc^ {2}$ one achieves to the following nonrelativistic
equation for $\varphi(\bf{x})$,
\begin{equation}
\left[\frac{p^{2}}{2m}+\frac{1}{2}m\omega^{2}r^{2}-\frac{3}{2}\hslash\omega-\frac{m\omega^{2}}{2\hbar}\boldsymbol{\theta}\cdot\mathbf{L}+\frac{m\omega^{2}}{8\hbar^{2}}(\boldsymbol{\theta}\times\mathbf{p})^{2}\right]\varphi=E\varphi.\label{eq:Klein2}\end{equation}
The first three terms into brackets contains the well-known Hamiltonian
of the nonrelativistic harmonic oscillator added by a constant term,
whereas the other two terms constitute the $\theta$-dependent Hamiltonian
$(\hat{H}^{\theta})$. It should be noted that the linear term in
$\theta$ is very similar to the interaction between the magnetic
field and the magnetic dipole moment associated with the orbital angular
momentum. The quadratic $\theta$-term can also be interpreted as
an electric dipole-dipole interaction, where $\boldsymbol{\mu}_{e}\propto\boldsymbol{\theta}\times\mathbf{p}$
\cite{Mirza,Riad}.

Our main objective is to evaluate the first-order corrections on the
energy spectrum yielded by $\hat{H}^{\theta}$, into the framework
of the perturbation theory. The first thing to do is to write the
exact eigenfunctions and the eigenvalues of the unperturbed Hamiltonian
$(\hat{H}_{ho})$ \cite{Strange}:
\begin{eqnarray}
&& \left[\frac{p^{2}}{2m}+\frac{1}{2}m\omega^{2}r^{2}-\frac{3}{2}\hslash\omega\right]\varphi_{nlm_{l}}  =\hbar\omega(2n+l)\varphi_{nlm_{l}} = E_{nl}\varphi_{nlm_{l}},\label{eq:Klein3}\\
&& \varphi_{nlm_{l}}(r,\theta,\phi) = R_{nl}(r)Y_{l m_{l}}(\theta,\phi),\label{eq:Klein4}\\
&& R_{nl}(r)  = \frac{A_{nl}}{r}\exp\left(-\frac{m\omega r^{2}}{2\hslash}\right)\left[\left(\frac{m\omega}{\hslash}\right)^{1/2}r\right]^{l+1}L_{n}^{l+1/2}(\frac{m\omega r^{2}}{\hslash}),\label{eq:Klein5}\\
&& A_{nl} =\left[\sqrt{\frac{m\omega}{\hslash\pi}}\frac{2^{n+l+2}n!}{(2n+2l+1)!!}\right]^{1/2},\label{eq:Klein6}\end{eqnarray}
 where $A_{nl}$ is the normalization constant, $Y_{l m_{l}}(\theta,\phi)$ are
standard spherical harmonics and $L_{n}^{l+1/2}(x)$ (with $n=0,1,2\ldots$)
are the associated Laguerre polynomials (see Ref. \cite{Gradshtein} for
definition). Hence, the stationary states $\varphi_{nlm_{l}}$ are also eigenstates of $\hat{\bf L}^{2}$ and $\hat{L}_{z}$.

It follows from Eq. (\ref{eq:Klein3}), that the energy only
depends on the quantum number $N=2n+l$ and the levels with $N\geq1$
are degenerate. Thus, according to degenerate perturbation theory,
it is necessary to diagonalize the matrix $\langle n^{'}l^{'}m_{l}^{'}\vert\hat{H}^{\theta}\vert nlm_{l}\rangle$
inside each of the degenerate subspaces of $\hat{H}_{ho}$. The first-order
energy corrections are the eigenvalues of this matrix. To be more
specific, we shall calculate the corrections to $N=0,1,2$.

First of all, let us note that the matrix element associated with
the term $\hat{H}_{1}^{\theta}=-(m\omega^{2}/2\hbar)\boldsymbol{\theta}\cdot\mathbf{L}$
is clearly diagonal and generates a Zeeman-like shift. For the case
of the noncommutative $\theta$-vector aligned along the $z$-axis: $\boldsymbol{\theta}=\theta_{z}\hat{\bf z}$ (which it is accomplished by a rotation or a redefinition of coordinates), one obtains
\begin{eqnarray}
\langle n^{'}l^{'}m_{l}^{'}\vert\hat{H}_{1}^{\theta}\vert nlm_{l}\rangle & = &-\frac{m\omega^{2}}{2\hslash}\langle n^{'}l^{'}m_{l}^{'}\vert\boldsymbol{\theta}\cdot\mathbf{L}\vert nlm_{l}\rangle\nonumber \\
 & = &-\frac{m\omega^{2}}{2\hslash}\int_{0}^{\infty}r^{2}R_{n^{'}l^{'}}(r)R_{nl}(r)dr\int_{0}^{4\pi}Y_{l^{'}m_{l}^{'}}^{*}\left[\theta_{z}\hat{L}_{z}\right]Y_{lm_{l}}d\Omega\nonumber \\
 & = &-\frac{m\omega^{2}\theta_{z}}{2}m_{l}\delta_{n^{'}n}\delta_{l^{'}l}\delta_{m_{l}^{'}m_{l}},\label{eq:Klein7}\end{eqnarray}
where we have taken into account the eigenvalue equation $\hat{L}_{z}Y_{lm_{l}}=\hslash m_{l}Y_{l m_{l}}$ and the orthogonality relation between
the eigenfunctions. The magnitude order of this correction is $m\omega^{2}\theta_{z}/2$. 

Moreover, the diagonal elements of $\hat{H}_{2}^{\theta}=m\omega^{2}(\boldsymbol{\theta}\times\mathbf{p})^{2}/8\hbar^{2}$
can be evaluated as following:
\begin{align}
\langle nlm_{l}\vert\hat{H}_{2}^{\theta}\vert nlm_{l}\rangle & =\frac{m\omega^{2}}{8\hbar^{2}}\langle nlm_{l}\vert(\boldsymbol{\theta}\times\mathbf{p})^{2}\vert nlm_{l}\rangle\nonumber \\
 & =\frac{m\omega^{2}}{8\hslash^{2}}\left[\langle nlm_{l}\vert\theta^{2}_{z}p^{2}\vert nlm_{l}\rangle-\langle nlm_{l}\vert(\boldsymbol{\theta}\cdot\mathbf{p})^{2}\vert nlm_{l}\rangle\right].\label{eq:Klein8}\end{align}
According to Eq. \eqref{eq:Klein3}, the first term in \eqref{eq:Klein8} can be written as
\begin{align}
\langle nlm_{l}\vert p^{2}\vert nlm_{l}\rangle & =\langle nlm_{l}\vert\left[2m\varepsilon_{nl}-m^{2}\omega^{2}r^{2}\right]\vert nlm_{l}\rangle\nonumber \\
 & =m\hbar\omega\left(2n+l+\frac{3}{2}\right),\label{eq:Klein9}\end{align} with $\varepsilon_{nl}=\hslash\omega(2n+l+3/2)$. We have used the well-known relation $\int_{0}^{\infty}dxe^{-x}x^{\alpha+1}\left[L_{n}^{\alpha}(x)\right]^{2}=\frac{\Gamma(n+\alpha+1)}{n!}(2n+\alpha+1)$. 

The second term in Eq. (\ref{eq:Klein8}) is a bit more complicated.
If we take into account that $\mathbf{p}=\frac{m}{i\hslash}[\mathbf{x},\frac{p^{2}}{2m}]$,
then
\begin{align}
\langle nlm_{l}\vert(\boldsymbol{\theta}\cdot\mathbf{p})^{2}\vert nlm_{l}\rangle & =-\frac{m^{2}}{\hbar^{2}}\langle nlm_{l}\vert[\boldsymbol{\theta}\cdot\mathbf{x},\frac{p^{2}}{2m}][\boldsymbol{\theta}\cdot\mathbf{x},\frac{p^{2}}{2m}]\vert nlm_{l}\rangle\nonumber \\
 & =\frac{m^{2}\theta_{z}^{2}}{\hbar^{2}}\!\!\!\sum_{\{n^{'},l^{'},m_{l}^{'}\}}\!\!\!\left(\varepsilon_{nl}-\varepsilon_{n^{'}l^{'}}\right)^{2}\left|\langle n^{'}l^{'}m_{l}^{'}\vert r\cos\theta\vert nlm_{l}\rangle\right|^{2},\end{align}
where we have applied the closure relation to the basis $\{\vert nlm_{l}\rangle\}$
between the commutators. So the angular integration is performed by means of result
\begin{eqnarray}
\langle l^{'}m_{l}^{'}\mid\cos\theta\mid lm_{l}\rangle= &&\left[\frac{(l-m_{l}+1)(l+m_{l}+1)}{(2l+1)(2l+3)}\right]^{1/2}\!\!\!\delta_{m^{\prime},m}\delta_{l^{\prime},l+1}\nonumber\\
&&+\left[\frac{(l-m_{l})(l+m_{l})}{(2l-1)(2l+1)}\right]^{1/2}\!\!\!\delta_{m^{\prime},m}\delta_{l^{\prime},l-1}.\end{eqnarray}
The remaining radial integration can be explicitly calculated by using
the recurrence relations for the associated Laguerre polynomials \cite{Gradshtein}:
$xL_{n}^{k+1}=(n+k+1)L_{n}^{k}-(n+1)L_{n+1}^{k}$ and $L_{n}^{k-1}=L_{n}^{k}-L_{n-1}^{k}$.

These results enable us to write the diagonal elements of $\hat{H}_{2}^{\theta}$
in the form:\begin{eqnarray}
\langle nlm_{l}\vert\hat{H}_{2}^{\theta}\vert nlm_{l}\rangle &=& \frac{m^{2}\omega^{3}\theta_{z}^{2}}{8\hslash}\left(2n+l+\frac{3}{2}\right)\times\nonumber \\
 && \!\!\!\left[1-\left(\frac{(l-m_{l}+1)(l+m_{l}+1)}{(2l+1)(2l+3)}+\frac{(l-m_{l})(l+m_{l})}{(2l-1)(2l+1)}\right)\right],\label{eq:Klein10}\end{eqnarray}
with multiplicative factor of strength $m^{2}\omega^{3}\theta_{z}^{2}/8\hbar$.
As a remarkable result, we have seen a factor of $\hbar$ in the denominator
to the earlier expression. On the other hand, it has been supposed in Ref. \cite{Bertolami} 
that the noncommutative length scale is of order $\theta\leq10^{-30}\ m^{2}$. In this manner, the two terms $\hat{H}_{1}^{\theta}$
($\theta$-linear) and $\hat{H}_{2}^{\theta}$ ($\theta$-quadratic)
must be treated on an equal footing. In summary, the total energy
shift is due the whole matrix $\mathbf{\hat{H}^{\theta}=}\langle n^{'}l^{'}m_{l}^{'}\vert\hat{H}_{1}^{\theta}+\hat{H}_{2}^{\theta}\vert nlm_{l}\rangle$. 

Furthermore, it is easy to see that $[\hat{H}^{\theta},\hat{L}_{z}]=0$,
but it does not occur with $\hat{\mathbf{L}}^{2}$. Consequently,
the perturbation can mix states with different values of orbital angular
momentum, but the matrix elements are non-zero only between states
with the same value of $m_{l}$.

Now, we must calculate the various matrix elements. For this goal, it is convenient to define:
\begin{equation}
\alpha\equiv m\omega^{2}\theta_{z},\ \ \ \ \beta\equiv\frac{m^{2}\omega^{3}\theta_{z}^{2}}{\hbar}.\label{eq:Klein11}\end{equation}
\begin{itemize}

\item $N=0$; $n=l=0$.

The ground state $(E_{N=0}=0)$ is non-degenerate; the first-order correction only shifts the energy as a whole by a quantity:\begin{equation}
\langle000\vert\hat{H}^{\theta}\vert000\rangle=\frac{\beta}{8}.\label{eq:Klein12}\end{equation}
\item $N=1$; $n=0,$ $l=1$ and $m_{l}=0,\pm1$.

The first excited state $(E_{N=1}=\hbar\omega)$ is three-fold degenerate. The $3\times3$ matrix representing $\hat{H}^{\theta}$ is diagonal:
\begin{equation}
\mathbf{\hat{H}^{\theta}}=\left(\begin{array}{ccc}
\frac{\alpha}{2}+\frac{\beta}{4} & 0 & 0\\
0 & \frac{\beta}{8} & 0\\
0 & 0 & -\frac{\alpha}{2}+\frac{\beta}{4}\end{array}\right).\end{equation}
\item $N=2$; $n=0,$ $l=2,$ $m_{l}=0,\pm2,\pm1,$ or $n=1,$
$l=m_{l}=0$.

The second excited state $(E_{N=2}=2\hbar\omega)$ is six-fold degenerate. The $6\times6$ matrix representing $\hat{H}^{\theta}$ can be written
(the basis vectors are arranged in the order $\vert0,2,-2\rangle$,
$\vert0,2,-1\rangle$, $\vert0,2,0\rangle$,$\vert0,2,1\rangle$,$\vert0,2,2\rangle$,$\vert1,0,0\rangle$):
\begin{equation}
\mathbf{\hat{H}^{\theta}}=\left(\begin{array}{cccccc}
\alpha+\frac{3\beta}{8} & 0 & 0 & 0 & 0 & 0\\
0 & \frac{\alpha}{2}+\frac{\beta}{4} & 0 & 0 & 0 & 0\\
0 & 0 & \frac{5\beta}{24} & 0 & 0 & \frac{\beta}{6\sqrt{2}}\\
0 & 0 & 0 & -\frac{\alpha}{2}+\frac{\beta}{4} & 0 & 0\\
0 & 0 & 0 & 0 & -\alpha+\frac{3\beta}{8} & 0\\
0 & 0 & \frac{\beta}{6\sqrt{2}} & 0 & 0 & \frac{7\beta}{24}\end{array}\right),\end{equation}
 whose eigenvalues are $\left\{ -\frac{\alpha}{2}+\frac{\beta}{4},\frac{\alpha}{2}+\frac{\beta}{4},-\alpha+\frac{3\beta}{8},\alpha+\frac{3\beta}{8},\frac{\beta}{8},\frac{3\beta}{8}\right\} $. 

\end{itemize}

The splitting of the energy levels of the Klein-Gordon oscillator are shown in Fig.~\ref{fig:Shift-energy-to-KleinGordon} as a function of the noncommutative $\theta$-parameter. 

As a result, we have observed that the noncommutative space effects,
closed by $\hat{H}^{\theta}$, yielded effective shifts on the Klein-Gordon
oscillator spectrum. This result indicates the complete breakdown
of the degeneracy, with the energy corrections depending on $n,$
$l,$ and $m_{l}$ quantum numbers. Further, if we take a vanishing
$\theta$, we retrieve the typical result of a commutative space.

It is worth mentioning that the Klein-Gordon oscillator in a noncommutative space admits an exact solution when we choose another basis of eigenfunctions to take advantage of the symmetry of the problem. In fact, one can rewrite Eq. (\ref{eq.2}) in Cartesian coordinates as ($\boldsymbol{\theta}=\theta_{z}\hat{\bf z}$):
\begin{equation}
H\left|\varphi\right\rangle =\left(H_{xy}+H_{z}\right)\left|\varphi\right\rangle =\left(W^{2}-m^{2}c^{4}\right)\left|\varphi\right\rangle  
,\label{eq21}\end{equation} where:
\begin{eqnarray}
H_{xy} & = & c^{2}\left[\left(1+\frac{m^{2}\omega^{2}\theta_{z}^{2}}{4\hslash^{2}}\right)(p_{x}^{2}+p_{y}^{2})+m^{2}\omega^{2}(x^{2}+y^{2})\right],\\
H_{z} & = & c^{2}\left[p_{z}^{2}+m^{2}\omega^{2}z^{2}-3m\hslash\omega-\frac{m^{2}\omega^{2}\theta_{z}}{\hslash}\hat{L}_{z}\right].
\end{eqnarray}To obtain the energy eigenvalues in Eq. (\ref{eq21}),    
we defined the operators $a_{\pm}$ and $a_{z}$ in the following form:
\begin{eqnarray}
a_{\pm} & = & \frac{1}{2}\left[\lambda\left(x\pm iy\right)+\frac{i}{\lambda\hslash}\left(p_{x}\pm ip_{y}\right)\right],\\
a_{z} & = & \frac{1}{\sqrt{2}}\left(\lambda z+\frac{i}{\lambda\hslash}p_{z}\right),
\end{eqnarray}
with $\lambda=\sqrt{m\omega/\hslash}$. Now, it is not difficult to see that $H$ can be expressed in terms of the number operators $N_{\pm}=a_{\pm}^{\dagger}a_{\pm}$ and $N_{z}=a^{\dagger}_{z}a_{z}$ as follows \cite{Mirza}:
\begin{eqnarray}
H_{xy} & = & 2mc^{2}\hslash\omega\sqrt{1+\frac{m^{2}\omega^{2}\theta_{z}^{2}}{4\hslash^{2}}}\left(N_{+}+N_{-}+1\right),\\
H_{z} & = & 2mc^{2}\hslash\omega\left(N_{z}+\frac{1}{2}\right)-3mc^{2}\hslash\omega-\frac{m^{2}c^{2}\omega^{2}\theta_{z}}{\hslash}\hslash\left(N_{-}-N_{+}\right).
\end{eqnarray}The common eigenvectors $\left|n_{+},n_{-},n_{z}\right\rangle$ of $H$ and $\hat{L}_{z}$ can be obtained by methods similar to the conventional harmonic oscillator. The relevant point here is that this exact result is compatible with
our perturbative analysis. Indeed, the exact energy levels corresponding
to Eq. (\ref{eq21}) can be explicitly written as
\begin{eqnarray}
W^{2}-m^{2}c^{4} & = & 2mc^{2}\left[  \hslash\omega \sqrt{1+\frac{m^{2}\omega^{2}\theta_{z}^{2}}{4\hslash^{2}}}\left(n_{+}+n_{-}+1\right)\right.\nonumber \\
 &  & \left.+\hslash\omega\left(n_{z}+\frac{1}{2}\right)-\frac{3}{2}\hslash\omega-\frac{m\omega^{2}\theta_{z}}{2}\left(n_{-}-n_{+}\right)\right]
\label{eq:exactEnergy},\end{eqnarray}
where $n_{+}$, $n_{-}$ and $n_{z}$ are positive integers or zero
associated with $N_{+}$, $N_{-}$ and $N_{z}$ respectively. A comparison
with the perturbation method can easily be made by taking the following
approximations:
\begin{equation}
W^{2}-m^{2}c^{4}\approxeq2mc^{2}E\ \ \mbox{and}\ \ \sqrt{1+\frac{m^{2}\omega^{2}\theta_{z}^{2}}{4\hslash^{2}}}\approxeq1+\frac{m^{2}\omega^{2}\theta_{z}^{2}}{8\hslash^{2}},
\end{equation}
which along with (\ref{eq:exactEnergy}) leads us to
\begin{equation}
E=\hslash\omega\left(n_{+}+n_{-}+n_{z}\right)-\frac{\alpha}{2}\left(n_{-}-n_{+}\right)+\frac{\beta}{8}\left(n_{+}+n_{-}+1\right),
\end{equation}
with $E$ being the nonrelativistic energy and $\alpha$, $\beta$
defined as in (\ref{eq:Klein11}). Now, we can identify $N=2n+l=n_{+}+n_{-}+n_{z}$
and as result, the energy levels are not degenerate and the corrections
induced by the noncommutativity are the same as those obtained earlier.
For example, if $N=0$ then automatically we have $n_{+}=n_{-}=n_{z}=0$
and $E=\beta/8$, in complete agreement with (\ref{eq:Klein12}).

\section{Energy Corrections of the Dirac Oscillator in a Noncommutative Space}

The relativistic wave equation for free fermions in 4-dimensional space-time
is the usual Dirac equation $(i\hbar\gamma^{\mu}\partial_{\mu}-mc)\psi=0$ 
\footnote{The Dirac matrices are written as: $\gamma^{0}=\beta=\left(\begin{array}{cc}
I & 0\\
0 & -I\end{array}\right)$, $\gamma^{i}=\beta\alpha^{i}=\left(\begin{array}{cc}
0 & \sigma^{i}\\
-\sigma^{i} & 0\end{array}\right)$, with $\sigma^{i}=(\sigma_{x},\sigma_{y},\sigma_{z})$ being the
usual Pauli matrices.%
}. In order to get the Dirac oscillator, we introduce an external potential
by a non-minimal coupling through the replacement $\mathbf{p}\rightarrow\mathbf{p}-im\omega\beta\mathbf{x}$
\cite{Mos}:
\begin{equation}
\left(c\boldsymbol{\alpha}\cdot\left(\mathbf{p}-im\omega\beta\mathbf{x}\right)+\beta mc^{2}\right)\psi(\mathbf{x})=W\psi(\mathbf{x}),\end{equation}
 where $\psi({\bf x},t)=\psi({\bf x})\exp(-iWt/\hslash)$.

As before, the Dirac oscillator equation in a noncommutative space is given by
\begin{equation}
\left[c\boldsymbol{\alpha}\cdot\left(\mathbf{p}-im\omega\beta\left(\mathbf{x}+\frac{\boldsymbol{\theta}\times\mathbf{p}}{2\hslash}\right)\right)+\beta mc^{2}\right]\psi=W\psi.\end{equation}

Following standard procedure, the equation for the upper component of $\psi=\left(\begin{array}{c}
\varphi\\
\chi\end{array}\right)$ can be written as:
\begin{eqnarray}
 && c^{2}\left[p^{2}+m^{2}\omega^{2}r^{2}-3m\hslash\omega-\frac{4m\omega}{\hslash}\mathbf{S}\cdot\mathbf{L}-\frac{m^{2}\omega^{2}}{\hslash}\boldsymbol{\theta}\cdot(\mathbf{L}+2\mathbf{S})\right.\nonumber \\
 && +\left.\frac{2m\omega}{\hslash^{2}}(\mathbf{S}\times\mathbf{p})\cdot(\boldsymbol{\theta}\times\mathbf{p})+\frac{m^{2}\omega^{2}}{4\hslash^{2}}(\boldsymbol{\theta}\times\mathbf{p})^{2}\right]\varphi=(W^{2}-m^{2}c^{4})\varphi,\label{eq:Dirac1}\end{eqnarray}
where $\mathbf{L}$ is the orbital angular momentum,  $\mathbf{S}=(\hslash/2)\boldsymbol{\sigma}$
is the spin and $r=\sqrt{\mathbf{x}\cdot\mathbf{x}}$. The above equation has no exact solution \cite{Mirza}, thus, a perturbative approach is needed.

As in the previous discussion, we are interested in the  nonrelativistic limit of (\ref{eq:Dirac1}). Here, this restriction implies
at the following nonrelativistic $\theta$-modified Hamiltonian for
the Dirac oscillator:\begin{eqnarray}
\hat{H} & = & \left\{ \left[\frac{p^{2}}{2m}+\frac{m\omega^{2}r^{2}}{2}-\frac{3\hslash\omega}{2}-\frac{2\omega}{\hslash}\mathbf{S}\cdot\mathbf{L}\right]\right.\nonumber \\
 &  & \!\!\left.-\frac{m\omega^{2}}{2\hslash}\boldsymbol{\theta}\cdot(\mathbf{L}+2\mathbf{S})+\frac{\omega}{\hslash^{2}}(\mathbf{S}\times\mathbf{p})\cdot(\boldsymbol{\theta}\times\mathbf{p})+\frac{m\omega^{2}}{8\hslash^{2}}(\boldsymbol{\theta}\times\mathbf{p})^{2}\right\} .\label{eq:Dirac2}\end{eqnarray}
 We see that the ordinary Hamiltonian of the Dirac oscillator $(\hat{H}_{DO})$
appears between brackets. The other terms compose the $\theta$-dependent
Hamiltonian $(\hat{H}^{\theta})$. Clearly, if the spin $\mathbf{S}$
is ignored, we recover the Eq. (\ref{eq:Klein2}) for the Klein-Gordon
oscillator. Moreover, the term associated with $\boldsymbol{\theta}\cdot(\mathbf{L}+2\mathbf{S})$
is very similar to the correction that leads to the anomalous Zeeman effect \cite{Banerjee}.

Our purpose is to determine the contribution of $\hat{H}^{\theta}$
on the energy spectrum of $\hat{H}_{DO}$. Since we have now the presence
of terms involving the spin operator, it is more suitable to work
with the eigenstates common of ${\bf L^{2}}$, ${\bf S}^{2}$, ${\bf J}^{2}$
and $J_{z}$, where $\mathbf{J=L+S}$ is the total angular momentum.
In particular, it is easy to verify that $[\hat{H}_{DO},{\bf J}]=0$.
In this way, the corresponding eigenfunctions of $\hat{H}_{DO}$ can
be split into $\psi_{nljm_{j}}=R_{nl}(r)\Omega_{jl}^{m_{j}}(\theta,\phi)$,
with $n$, $l$, $j$, $m_{j}$ being the associated quantum numbers.
The radial components are the same that in (\ref{eq:Klein5}), with
the angular part of the wave function being given by \begin{equation}
\Omega_{j=l\pm\frac{1}{2},l}^{m_{j}}=\left(\begin{array}{c}
\pm\sqrt{\frac{l\pm m_{j}+\frac{1}{2}}{2l+1}}Y_{l}^{m_{j}-\frac{1}{2}}(\theta,\phi)\\
\sqrt{\frac{l\mp m_{j}+\frac{1}{2}}{2l+1}}Y_{l}^{m_{j}+\frac{1}{2}}(\theta,\phi)\end{array}\right),\label{eq:Dirac3}\end{equation}
where $Y_{l}^{m_{j}\pm\frac{1}{2}}$ is the spherical harmonic function,
with $l\geq0$, $j=l\pm\frac{1}{2}$ and $-j\leq m_{j}\leq j$.

The energy spectrum of $\hat{H}_{DO}$ can be written as \cite{Strange}:
\begin{equation}
E=\left\{ \begin{array}{lll}
(N-j+1/2)\hbar\omega=2n\hbar\omega & \ \mbox{if}\  & j=l+\frac{1}{2}\\
(N+j+3/2)\hbar\omega=(2n+2l+1)\hbar\omega & \ \mbox{if}\  & j=l-\frac{1}{2}\end{array}\right..\label{eq:Dirac4}\end{equation}

It should be noted the remarkable amount of degeneracy found in the
previous expression. For $j=l+\frac{1}{2}$, the energy depends only
the values of $n$. Since $l$ is any positive integer or zero, the
degeneracy of this energy level is infinite. For $j=l-\frac{1}{2}$,
the energy depends on the sum $k=n+l$ with $k\geq1$, but now the
degeneracy remains finite, increasing with the $k$ value. Furthermore,
if $j=\frac{1}{2}$ $(l=0)$ the energy value is the same that in
(\ref{eq:Klein3}) and all states are two-fold degenerate (with fixed
$n$). When compared with the nonrelativistic Klein-Gordon oscillator,
the previous analysis shows the non-trivial effect induced by spin-orbit
coupling on the energy levels of the system.

Finally, as in the Sec. 2, we shall calculate the $\theta$-modifications
on the energy levels by determining the eigenvalues of the matrix
$\mathbf{\hat{H}^{\theta}=}\langle n^{'}l^{'}j^{'}m_{j}^{'}\vert\hat{H}^{\theta}\vert nljm_{j}\rangle$
where \begin{equation}
\hat{H}^{\theta}=-\frac{m\omega^{2}}{2\hslash}\boldsymbol{\theta}\cdot(\mathbf{L}+2\mathbf{S})+\frac{\omega}{\hslash^{2}}(\mathbf{S}\times\mathbf{p})\cdot(\boldsymbol{\theta}\times\mathbf{p})+\frac{m\omega^{2}}{8\hslash^{2}}(\boldsymbol{\theta}\times\mathbf{p})^{2},\end{equation}
with non-vanishing elements only when $m_{j}^{'}=m_{j}$ (it is not
difficult to see that now $[\hat{J}_{z},\hat{H^{\theta}}]=0$). 

To do so, we confine ourselves to the case where $j=l-\frac{1}{2}$ $(l\neq0)$
and $k=n+l=1,2,3$. The case $j=l+\frac{1}{2}$ is more complicated
because in principle, we have to diagonalize an infinite matrix. Thus,
we obtain:
\begin{itemize}

\item $k=1$; $n=0,$ $l=1,$ $j=\frac{1}{2}$ and $m_{j}=\pm\frac{1}{2}$.

This energy level $(E_{k=1}=3\hbar\omega)$ is two-fold degenerate (essential degeneracy). The $2\times2$ matrix representing $\hat{H}^{\theta}$
is diagonal:
\begin{equation}
\hat{{\bf H}}^{\theta}=\left(\begin{array}{cc}
\alpha+\frac{5\beta}{24} & 0\\
0 & -\alpha+\frac{5\beta}{24}\end{array}\right),\end{equation}
where $\alpha$ and $\beta$ are defined as in (\ref{eq:Klein11}). 

\item $k=2$; $n=0,$ $l=2$, $j=\frac{3}{2}$ and $m_{j}=\pm\frac{3}{2},\pm\frac{1}{2}$
or $n=1,$ $l=1,$ $j=\frac{1}{2}$ and $m_{j}=\pm\frac{1}{2}$.

This energy level $(E_{k=2}=5\hbar\omega)$ is six-fold degenerate
(essential and accidental degeneracies). The $6\times6$ matrix representing
$\hat{H}^{\theta}$ is diagonal too:
\begin{equation}
\hat{{\bf H}}^{\theta}=\left(\begin{array}{cccccc}
2\alpha+\frac{7\beta}{20} & 0 & 0 & 0 & 0 & 0\\
0 & \frac{2\alpha}{3}+\frac{7\beta}{30} & 0 & 0 & 0 & 0\\
0 & 0 & -\frac{2\alpha}{3}+\frac{7\beta}{30} & 0 & 0 & 0\\
0 & 0 & 0 & -2\alpha+\frac{7\beta}{20} & 0 & 0\\
0 & 0 & 0 & 0 & \frac{5\alpha}{3}+\frac{3\beta}{8} & 0\\
0 & 0 & 0 & 0 & 0 & -\frac{5\alpha}{3}+\frac{3\beta}{8}\end{array}\right).\end{equation}

\item $k=3$; $n=0,$ $l=3,$ $j=\frac{5}{2}$ and $m_{j}=\pm\frac{5}{2},\pm\frac{3}{2},\pm\frac{1}{2}$
or $n=1,$ $l=2,$ $j=\frac{3}{2}$ and $m_{j}=\pm\frac{3}{2},\pm\frac{1}{2}$
or $n=2$, $l=1,$ $j=\frac{1}{2}$ and $m_{j}=\pm\frac{1}{2}$.

This energy level $(E_{k=3}=7\hbar\omega)$ is twelve-fold degenerate
and the $12\times12$ matrix representing $\hat{H}^{\theta}$ is non-diagonal.
Taking into account the eigenfunctions of $\hat{H}_{DO}$, it is possible
to show that the eigenvalues of $\hat{H}^{\theta}$ are all non-degenerate
and have the form: 
\begin{eqnarray}
&&\left\{-\frac{9\alpha}{5}+\frac{99\beta}{280},\frac{9\alpha}{5}+\frac{99\beta}{280},-\frac{14\alpha}{15}+\frac{11\beta}{30},\frac{14\alpha}{15}+\frac{11\beta}{30},\right.\nonumber\\
&&\ \ -3\alpha+\frac{27\beta}{56},3\alpha+\frac{27\beta}{56},-\frac{14\alpha}{5}+\frac{11\beta}{20},\frac{14\alpha}{5}+\frac{11\beta}{20},\nonumber \\
&&\frac{1}{840}\left(1232\alpha+349\beta-2\sqrt{132496\alpha^{2}+38584\alpha\beta+3103\beta^{2}}\right),\nonumber\\
&&\frac{1}{840}\left(1232\alpha+349\beta+2\sqrt{132496\alpha^{2}+38584\alpha\beta+3103\beta^{2}}\right),\nonumber \\
&&\frac{1}{840}\left(-1232\alpha+349\beta-2\sqrt{132496\alpha^{2}-38584\alpha\beta+3103\beta^{2}}\right),\nonumber\\
&&\left.\frac{1}{840}\left(-1232\alpha+349\beta+2\sqrt{132496\alpha^{2}-38584\alpha\beta+3103\beta^{2}}\right)\right\}.
\end{eqnarray}
\end{itemize}

\begin{figure}[p]
\begin{minipage}[t]{0.49\linewidth}
\includegraphics[width=0.95\textwidth]{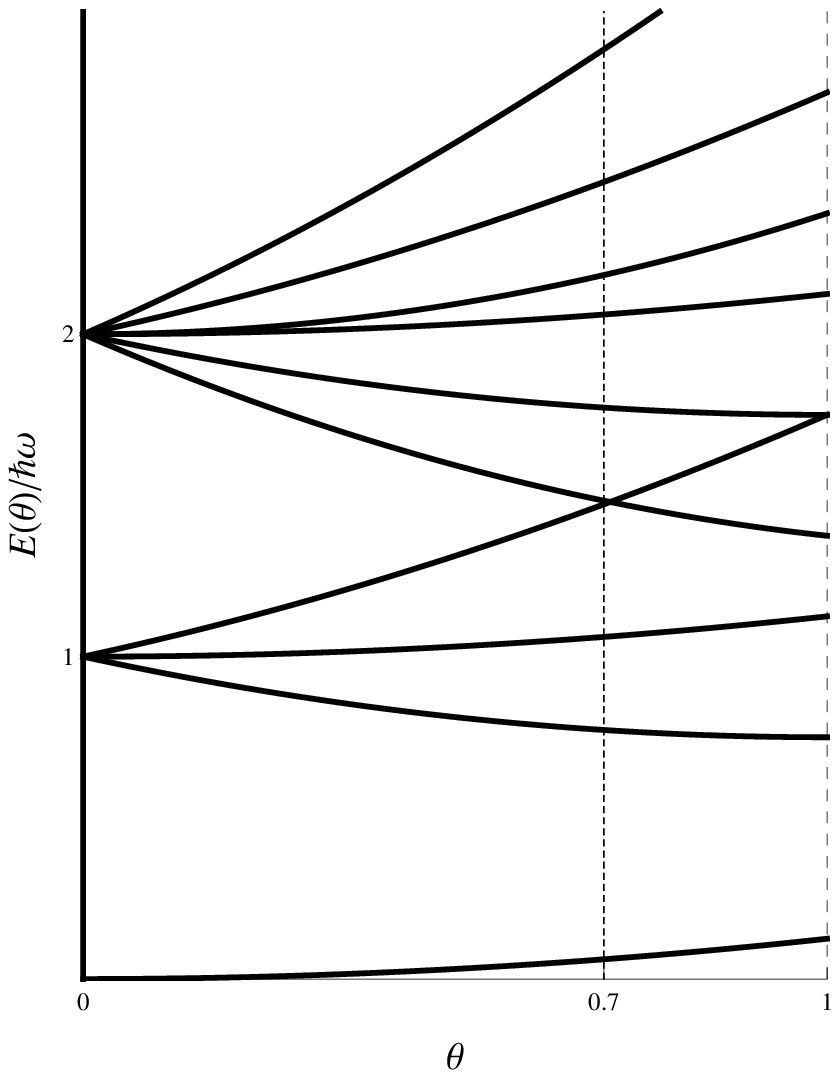}
\caption{Energy shift to the Klein-Gordon oscillator. When $\theta\geq0.7$, some additional degeneracies appear. It is
assumed that $m\omega/\hbar=1$.}
\label{fig:Shift-energy-to-KleinGordon}
\end{minipage}\hfill
\begin{minipage}[t]{0.49\linewidth}
\includegraphics[width=0.95\textwidth]{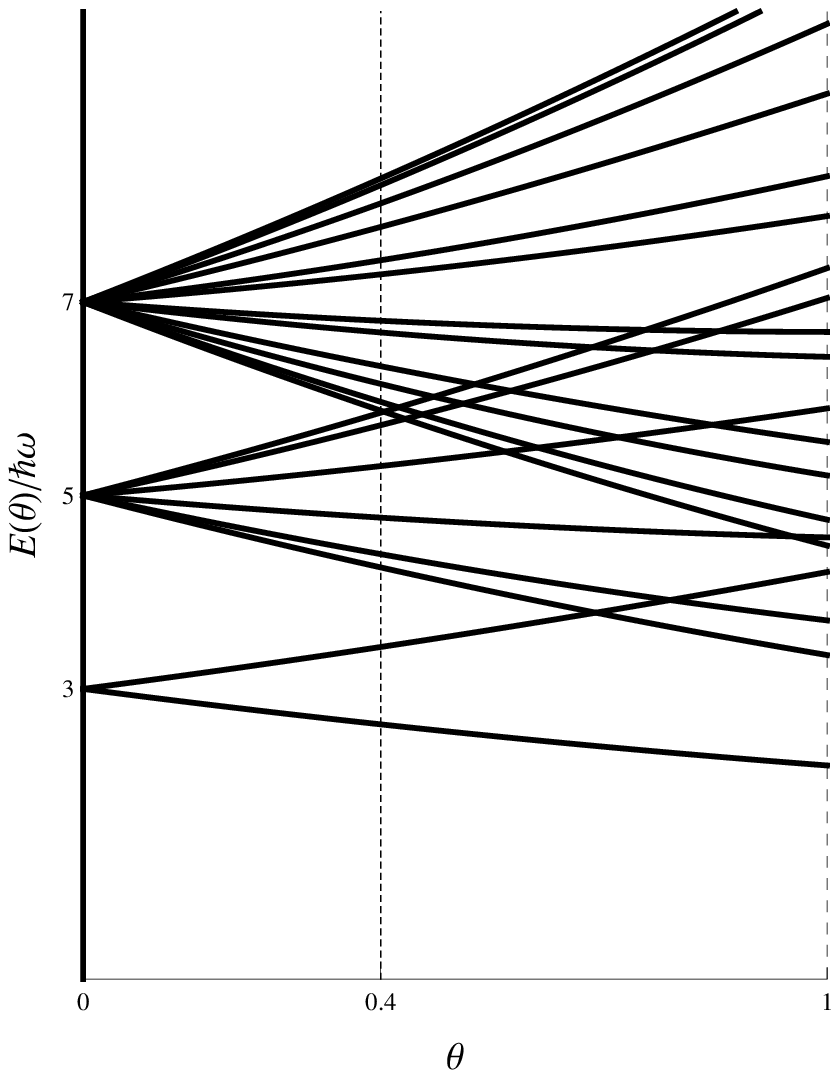}
\caption{Energy shift to the Dirac oscillator. When $\theta\geq0.4$, some additional degeneracies appear. It is assumed that $m\omega/\hbar=1$.}
\label{fig:Shift-energy-to-Dirac}
\end{minipage}\hfill
\end{figure}

Such as in the nonrelativistic Klein-Gordon oscillator, the spatial
noncommutativity is able to modify the fine structure of the spectrum,
with total lifting of the degeneracy on the energy levels considered.
It is important to point out here the difference between these results
and those reported in Ref. \cite{Thiago}, for nonrelativistic hydrogen
atom. In the latter, the degeneracy is only partially removed by the
noncommutativity. The corresponding splits of the energy levels to the Dirac oscillator
are shown in Fig.~\ref{fig:Shift-energy-to-Dirac} as a function of
the noncommutative $\theta$-parameter.

\section{Conclusions}

In this paper, we have studied the effects of spatial noncommutativity
on the energy spectrum of the Klein-Gordon and Dirac oscillators.
Indeed, the nonrelativistic limit has been worked out and the $\theta$-modified
Hamiltonians (derived from the Bopp shift) were determined. In both
systems, the first-order corrections induced by spatial noncommutativity
were able to completely remove the degeneracy of the energy levels
analyzed. In the case of the Dirac oscillator, we observed the presence
of terms depending on the spin operator and the noncommutative $\theta$-parameter, implying similar modifications to the anomalous Zeeman effect. It was also found that if the limit $\theta\rightarrow0$ is taken, then
we recover the results of the commutative case. Once that the Dirac
oscillator has been extensively explored in the literature (see Ref. \cite{Sadumi}
for a review on the subject), we expect that the above results can
be used to set up new bounds to the $\theta$-parameter magnitude.

Finally, we would like to point out that the same calculations are
possible for the noncommutative Kemmer Oscillator \cite{Zare}. Moreover,
in a recent paper the spin noncommutativity (different from the canonical)
has been proposed \cite{Kupriyanov}. Thus, an extension of our work
in this new context would be interesting.

\section*{Acknowledgments}
This work was partially supported by Conselho Nacional de Desenvolvimento
Científico e Tecnológico (CNPq). The author thanks M. Gomes and A. J. da Silva for useful discussions and comments. The author also thanks the reviewer for the positive suggestions.

\end{document}